\documentclass[aps,prl,showpacs,floatfix,twocolumn,amsmath,amssymb,dvips]{revtex4-1}%
\usepackage{amsmath}
\usepackage{epsfig}
\usepackage{graphicx}
\usepackage{dcolumn}
\usepackage{bm}
\usepackage{amsfonts}
\usepackage{amssymb}%
\usepackage[colorlinks=true,dvipdfm]{hyperref}

\begin{document}

\title{Renormalization-group theory for cooling first-order phase transitions in Potts models}
\author{Ning Liang and Fan Zhong}
\thanks{Corresponding author. E-mail: stszf@mail.sysu.edu.cn}
\affiliation{State Key Laboratory of Optoelectronic Materials and Technologies, School of
Physics, Sun Yat-sen University, Guangzhou 510275, People's
Republic of China}
\date{\today }

\begin{abstract}
We develop a dynamic field-theoretic renormalization-group (RG) theory for the cooling first-order phase transitions in the Potts model. It is suggested that the well-known imaginary fixed points of the $q$-state Potts model for $q>10/3$ in the RG theory are the origin of the dynamic scaling found recently, apart from the logarithmic corrections. This indicates that the real and imaginary fixed points of the Potts model are both physical and control the scalings of the continuous and discontinuous phase transitions, respectively, of the model. Our one-loop results for the scaling exponents are already not far away from the numerical results. Further, the scaling exponents depend on $q$ slightly only, in consistence with the numerical results. Therefore, the theory is believed to provide a natural explanation of the dynamic scaling including the scaling exponents and their scaling laws for various observables in the Potts model.
\end{abstract}

\maketitle


Recent experimental advances in manipulating real-time evolution of ultracold atoms~\cite{Greiner,Kinoshita,Hofferberth,Zhang} have stimulated a resurgence in studying the dynamics of continuous phase transitions. When a system is linearly cooled through a continue phase transition into a symmetry-broken ordered phase, the nonequilibrium scaling of topological defects formed by the Kibble-Zurek mechanism has been proposed~\cite{Kibble1,Kibble2,Zurek1,Zurek2} and tested in many classical and quantum systems~\cite{Dziarmaga,polrmp,inexper4}. In analogy to the well-known finite-size scaling (FSS) in which the correlation length of a system can be longer than its size in the vicinity of its critical point, a linear driving, the generalization of the cooling, introduces a finite time scale that is inevitably shorter than the correlation time near the critical point because of critical slowing down and thus finite-time scaling (FTS) of any observable besides the defect density has also been proposed~\cite{Zhong1,Zhong2} and tested~\cite{Chandran,Yin,Yin2,Liu,Huang,Huang2,Pelissetto}. Extension of FTS to an arbitrary driving has also led to a theory for a series of driven nonequilibrium critical phenomena resulting from the competitions of various involved time scales and initial conditions~\cite{Feng}.

Recently, the driven nonequilibrium approach was applied to first-order phase transitions (FOPTs)~\cite{Pana,Pelissetto,Pelissetto16}. By combining FTS and the FSS of FOPTs near their equilibrium transition points~\cite{Niehuis,FisherB,Privman,FisherP,Challa,Borg,Campostrini} with the exponential tunneling time between the two phases due to their interfaces~\cite{Berg}, a dynamic scaling near a temperature other than the equilibrium transition point was found with logarithmic corrections for the cooling FOPTs in the two-dimensional (2D) $q$-state Potts model with $q=10$ and $20$~\cite{Pelissetto16}. This spinodal-like scaling in systems with only short-range interactions was suggested to arise from the interplay between the exponential tunneling time and the droplet formation in the low-temperature phase. However, the origin of the scaling exponent and the scaling laws for different observables have yet to be identified~\cite{Pelissetto16}.

On the other hand, dynamic scalings of magnetic hysteresis~\cite{Rao,Lo,Tome,Jung,Sengupta,Dhar,He,Dhar93,Luse,Somoza,Ackaryya,cha} and of FOPTs ~\cite{zhang86,zhong94,zhonge1,zhonge2,zhongl,zhang95,zhongssc,liu,Kuang,zhongr,Yildiz,zhongl05,Fan} under driving were found some time ago and renormalization-group (RG) theories for the latter have been put forth~\cite{zhongl,zhongr,zhongl05,zhonge12,li12,zhong16,liang}. Without driving, scaling near weak FOPTs has been detected using a short-time relaxation approach~\cite{stZhengb1,Zhengb5}. In addition, dynamic scaling at the late stage of growth after sudden quenches to the two-phase region in phase ordering kinetics was once intensively studied~\cite{Bray,Gunton,Binder1,Binder2}. In mean-field theories that become exact when the interactions of the considered system are sufficiently long range, dynamic scaling near the spinodals can be worked out~\cite{Jung,Luse}. However, it is generally believed that sharply defined spinodals do not exist for systems with short-range interactions in contrary to the mean-field case~\cite{Gunton,Binder1,Binder2}. Nevertheless, it is also generally believed that there exists a crossover region that distinguishes the apparently different dynamics of nucleation and growth from spinodal decomposition at least at the early stage of an FOPT for systems with short-range interactions~\cite{Gunton,Binder1,Binder2}. One may argue that this crossover region is characterized by an instability point that is the correspondence of the spinodal point in the mean field theory and is shifted from the latter by fluctuations. An expansion near such instability points of a usual $\phi^4$ theory below its critical point then results in a $\phi^3$ theory for field-driven FOPTs~\cite{zhongl05,zhong16}. Although the fixed points of such a theory are imaginary in values, they are physical counter-intuitively in order for the $\phi^{3}$ theory to be mathematically convergent, as at the instability points, the system flows to the instability fixed points upon coarse graining~\cite{zhonge12}. Consequently, the instability fixed points are suggested to be the origin of the observed dynamic scaling~\cite{zhongl05,zhong16}. A similar RG theory for driven thermal FOPTs has also been developed~\cite{liang}. However, no clear evidence of an overall power-law relationship was found for the magnetic hysteresis in a sinusoidally oscillating field in 2D~\cite{Thomas,Sides98,Sides99}. The results of the Potts model now show that spinodal-like dynamic scaling does exist for FOPTs in systems with short-range interactions if logarithmic corrections are properly considered~\cite{Pelissetto16}.

Here we shall develop a dynamic field-theoretic RG theory for the cooling FOPTs in the Potts model. In accordance with the $\phi^3$ theory, we suggest that the well-known imaginary fixed points of the Potts model for $q>10/3$~\cite{Amit} are the origin of the dynamic scaling found in Ref.~\cite{Pelissetto16}, apart from the logarithmic corrections. Therefore, the real and imaginary fixed points of the Potts model are both physical and control the scalings of the continuous and discontinuous phase transitions, respectively. Our one-loop results for the hysteresis scaling exponents are already not far away from the numerical results. Further, we find that the difference of the hysteresis scaling exponents in $q=10$ and $20$ is only small, also in consistence with the numerical results, although the exponents do depend on the state number $q$. Therefore, we believe the theory provides a natural explanation of the dynamic scaling including the scaling exponents and their scaling laws for various observables in the Potts model.

The $q$-state Potts model~\cite{Potts,Wu} can be described by a field theory with the free-energy functional~\cite{Zia}
\begin{eqnarray}
{\cal F}[\phi _{i}]=&&\int d\mathrm{\mathbf{x}}\left\{ \frac{1}{2}r_0\phi _{i}\phi
_{i}+\frac{1}{2}\nabla \phi _{i}\nabla \phi _{i}+\frac{1}{3!}w_0Q_{ijk}{\phi }%
_{i}{\phi }_{j}{\phi }_{k}\right.\nonumber\\
&&\qquad\quad~\left.+\frac{1}{4!}g_0T_{ijkl}{\phi }_{i}{\phi }_{j}{\phi }%
_{k}{\phi }_{l}\right\},  \label{f1}
\end{eqnarray}
where $\phi_{i}$ is a $(q-1)$-dimensional order parameter field, $r_0$ is the reduced temperature, and $g_0$ and $w_0$ are coupling constants. In Eq.~(\ref{f1}), the tensors $Q_{ijk}$ and $T_{ijkl}$ are given by
\begin{equation}
Q_{ikl}=\sum_{\alpha =1}^{q}e_{i}^{\alpha
}e_{j}^{\alpha }e_{k}^{\alpha },\qquad T_{ijkl}=\sum_{\alpha =1}^{q}e_{i}^{\alpha }e_{j}^{\alpha }e_{k}^{\alpha
}e_{l}^{\alpha }.  \label{tensor1}
\end{equation}
via a set of $q$ vectors $e_{i}^{\alpha}$ in a $(q-1)$-dimensional space satisfying
\begin{equation}
e_{i}^{\alpha }e_{i}^{\beta }=q\delta ^{\alpha \beta}-1,\qquad
e_{i}^{\alpha
}e_{j}^{\alpha } =q\delta _{ij},\qquad
\sum_{\alpha =1}^{q}e_{i}^{\alpha } =0, \label{tensor}
\end{equation}
where $\delta$ is the Kronecker delta function. Summation over repeated indices is implied throughout.

The model~(\ref{f1}) has been rather well studied~\cite{Zia,Amit,Alcantara}. One notices that Eq.~(\ref{f1}) has a cubic term. As a result, its mean-field approximation exhibits an FOPT at $r_0=0$. However, for low $q$ and in low spatial dimensions, strong fluctuations change the transitions into continuous ones. Again owing to the cubic term, the quartic term can be ignored and the upper critical dimension of the model is six rather than four. RG theories for the model near $d=6$ dimensions via the $\epsilon=6-d$ expansions then shows that real fixed points exist and hence the transitions are continuous for $q<10/3$. This is consistent with the exact results that continuous transitions occur for $q\leq4$ in 2D~\cite{Baxter73,Wu}, though the transition for $q=3$ in 3D is now believed to be first order~\cite{Wu,Hartmann}. By contrast, for $q>10/3$, the fixed points are imaginary, which is interpreted as no existing accessible fixed points and the transitions are then first order.

On the basis of the $\phi^3$ theory~\cite{zhongl05,zhonge12,zhong16}, we suggest that the imaginary fixed points of the Potts model for FOPTs do not mean that these fixed points are not physical and thus are not accessible. Rather, the imaginariness of the fixed points is the prerequisite for their being physical, because the systems now become unstable near their instability points and thus analytical continuations are necessary.

In the following, we shall study the imaginary fixed points in dynamics that is required for the system to enter metastable states.

As usual, the nonconserved dynamics is governed by~\cite{Hohenberg,Folk}
\begin{equation}
\frac{\partial{{\phi}_{i}}}{\partial t}=-\lambda_0\frac{\delta {\cal F}[\phi_{i}]}{%
\delta{{\phi}_{i}}}+\zeta_{i}  \label{lang2}
\end{equation}
with the Gaussian white noise $\zeta_{i}$ obeying
\begin{align}
\langle\zeta_{i}(\mathbf{r},t)\rangle & =0,  \notag \\
\langle\zeta_{i}(\mathbf{r},t)\zeta_{j}(\mathbf{r^{\prime}}%
,t^{\prime})\rangle & =2\lambda_0 \delta(\mathbf{r}-\mathbf{r^{\prime}}%
)\delta(t-t^{\prime })\delta_{ij},  \label{noice 2}
\end{align}
where $\lambda_0$ is a kinetic coefficient. This model can be cast into a dynamic field theory with an action~\cite{Janssen79,janssen,Tauber,Justin,Vasilev}
\begin{eqnarray}
I[\phi _{i},{\tilde{\phi}}_{j}]=&&\int d\mathrm{\mathbf{x}}dt\left\{
\tilde{\phi}_{j}\left[ \delta _{ij}\left( \dot{\phi}_{i}{-\lambda_0 }%
r_0 {\phi }_{i}+{\lambda_0 \nabla ^{2}\phi }_{i}\right)\right.\right. \nonumber\\
&&\qquad\qquad\left.\left.+\frac{1}{2!}%
\lambda_0 w_0Q_{ijk}\phi _{i}\phi _{k}\right] -\lambda_0 {\tilde{\phi}}%
_{j}^{2}\right\},  \label{at1P}
\end{eqnarray}
where ${\tilde{\phi}}_i$ is a response field~\cite{martin}.

Driven thermal FOPTs for scalar models fall into three universality classes, although they are all controlled by the same imaginary fixed points~\cite{liang}. This ramification stems from the symmetry of the $\phi^3$ theory~\cite{Breuer}. A displacement of the order parameter mixes the coefficients of the linear and the quadratic terms of the free-energy functional. As a consequence, the driven thermal transition can exhibit either the field-like or the partial thermal classes, when the linear or the quadratic terms, respectively, are dominated. Moreover, crossover between the two classes can also occur~\cite{liang}. The third class is the purely thermal class for cooling transitions in the mean field models. Here, only the coefficient of the quadratic term, the reduced temperature, is varied and no external field, the linear term, is present. However, when fluctuations are taken into account, a linear term is always generated and the first two classes are recovered.

However, the continuous Potts model has a unique feature, which is expected to be shared by the lattice model due to universality. It is apparent that the one-loop contribution to the linear term is proportional to $Q_{ijj}$, which is equal to zero using Eqs.~(\ref{tensor1}) and (\ref{tensor}). Therefore, the linear term vanishes in the one-loop order, though the higher-order results do not. Therefore, the cooling FOPTs of the Potts model belong to the purely thermal class that is described by the purely massive theory, Eq.~(\ref{at1P}). By contrast, the heating transitions fall into the other two classes, because the expansion near the instability points always generate both the linear and the quadratic terms~\cite{liang}.

Within the RG theory for the universal scaling behavior of the model~\cite{Janssen79,janssen,Tauber,Justin,Vasilev}, one then needs to find the renormalization factors, defined as,
\begin{eqnarray}
\begin{array}{ll}
\phi _{i}\to\phi _{i0} =Z_{\phi }^{1/2}\phi _{i},&\widetilde{\phi }%
_{i}\to \widetilde{\phi }_{i0}=Z_{\widetilde{\phi }}^{1/2}\widetilde{\phi }_{i},\\
\lambda_0=Z_{\lambda }\lambda=Z_{{\phi }} ^{1/2}Z_{\widetilde{\phi }%
}^{-1/2}\lambda, &r_0=r_{s}+Z_{{\phi }}^{-1}Z_{\tau}\tau,\\
u_0 =Z_{w}u=w_0N_{d}^{1/2}\mu ^{-\epsilon /2},&\label{renormalization factorsP}
\end{array}
\end{eqnarray}
that cancel the ultraviolet divergences of the theory, where $0$ denotes the bare parameters, $\mu$ is an arbitrary momentum scale, $r_s$ results from the fluctuation shift of the transition temperature, and $N_{d}=2/[(4\pi)^{d/2}\Gamma(d/2)]$ with $\Gamma$ being the Euler gamma function. Upon employing the minimal RG scheme with dimensional regulation~\cite{hooft,Dominicis}, the renormalization factors to the one-loop order are
\begin{equation}
\begin{array}{ll}
Z_{{\phi }}=1-\alpha _{1}u^{2}/6\epsilon ,&
Z_{\widetilde{{\phi }}}=1-\alpha _{1}u^{2}/3\epsilon , \\
Z_{w}=1+(\alpha _{1}/4-\beta _{1})u^{3}/\epsilon,~ &
Z_{\tau }=1-3\alpha _{1}u^{3}/4\epsilon .
\end{array}\label{ZZP}
\end{equation}%
where $\alpha_1$ and $\beta_1$ are defined as~\cite{Amit}
\begin{eqnarray}
{Q_{ikl}Q_{jkl}}& =&{\alpha }_{1}{\delta }_{ij}=q^{2}(q-2)\delta_{ij},  \nonumber \\
Q_{ilm}Q_{jmn}Q_{knl}& =&{\beta }_{1}Q_{ijk}=q^{2}(q-3)Q_{ijk}.  \label{qcontract}
\end{eqnarray}
The static $Z$ factors agree with the extant ones~\cite{Amit,Alcantara}, and all factors in Eq.~(\ref{ZZP}) reduce to those of the scalar model in which case $\alpha _{1}=\beta_1=1$~\cite{zhongl05,zhong16}. Moreover, in the same limit, $Z_w=Z_{\tau}$, indicating that only three independent factors are required~\cite{liang}. However, for the Potts model, there are four independent factors similar to the usual $\phi^4$ universality class of critical phenomena, since no shift symmetry exists. We shall see shortly that the exponent of the order parameter $\beta$ is then different from $1$ in this case.

The scaling behavior can then be derived from the RG equation. In a cooling transition without the presence of a symmetry breaking external field, the averaged order parameter for any component may not be a good observable. Some of its variants may even behave distinctively~\cite{Huang}. Thus, we consider the connected two-point correlation function $G(\mathbf{k})$ at a momentum $\mathbf{k}$. From the scale independence of the renormalized functions, the RG equation for $G$ is~\cite{Tauber,Justin,Vasilev,zhongl05,zhong16}
\begin{equation}
\left( \mu \frac{\partial }{\partial \mu }+\beta \frac{\partial }{\partial
u}+\gamma _{\lambda }\lambda \frac{\partial }{\partial \lambda }%
+\gamma _{\tau }\tau \frac{\partial }{\partial \tau }+%
\gamma_{\phi} \right) G=0,  \label{RG6sp}
\end{equation}
where the Wilson functions are defined as
\begin{eqnarray}
\gamma _{\tau }(u)&=&\frac{\partial \ln Z_{\tau }}{%
\partial \ln\mu },\quad \widetilde{\gamma }(u)=\frac{\partial
\ln Z_{\widetilde{\phi }}}{\partial \ln\mu },\quad \beta (u)= \frac{\partial u}{\partial \ln\mu },\nonumber\\
\gamma_{\phi} (u)&=&\frac{\partial \ln Z_{\phi }}{\partial \ln\mu },\quad\gamma _{\lambda }(u)= \frac{\partial \ln \lambda}{\partial \ln\mu }=\frac{1}{2}\widetilde{\gamma }-\frac{1}{2}\gamma_{\phi}
\label{WilsonsPP}
\end{eqnarray}
at constant bare parameters. At the fixed point at which $\beta(u^*)=0$,
by combining with na\"{\i}ve dimensional analysis~\cite{Justin,Vasilev,Tauber}, the solution to Eq.~(\ref{RG6sp}) is~\cite{Tauber,Justin,Vasilev,zhongl05,zhong16}
\begin{equation}
G(\mathbf{k},\lambda ,\tau,u^{\ast })=\kappa ^{2\beta /\nu
}G(\mathbf{k}\kappa^{-1},\lambda t\kappa ^{z},\tau\kappa ^{-1/\nu },u^{\ast
}), \label{form4p}
\end{equation}
with the instability exponents defined as
\begin{equation}
\eta =\gamma ^{\ast }_{\phi},~~ \beta /\nu =(d-2+\eta )/2,~~ z=\ 2+\gamma
_{\lambda }^{\ast },~~ 1/\nu =2-\gamma _{\tau }^{\ast },
\label{Pexponents}
\end{equation}
which are the exact counterparts of the critical exponents, where $\kappa$ is a momentum scale.

From Eqs.~(\ref{ZZP}) and (\ref{WilsonsPP}), the fixed points are given by
\begin{equation}
\beta(u)=-\frac{1}{2}\epsilon u+\left(  \frac{1}{4}\alpha_1-\beta_1\right)
u^{3}=0.\label{POTTSFIX}%
\end{equation}
The nonzero solutions of the fixed points are
\begin{equation}
u^{\ast2}=\frac{2\epsilon}{(\alpha_{1}-4\beta_{1})}=\frac{2\epsilon}%
{q^{2}(10-3q)},\label{POTTSFIX1}%
\end{equation}
which are infrared stable for $\epsilon>0$ or $d<6$~\cite{zhong16}. However, in these spatial dimensions, real fixed points only exist for $q<10/3$ as pointed out above; and the imaginary ones correspond to FOPTs. Nevertheless, again as pointed out above, these imaginary fixed points are physical and thus accessible~\cite{zhonge12}. They control the scaling behavior for the FOPTs just as the real fixed points do for the continuous phase transitions.

By Eqs.~(\ref{ZZP}), (\ref{WilsonsPP}), (\ref{Pexponents}) and (\ref{POTTSFIX1}), the instability exponents of the imaginary fixed points are
\begin{eqnarray}
\eta  &=&\frac{\alpha _{1}\epsilon }{3(\alpha _{1}-4\beta _{1})},\qquad\qquad z=2+%
\frac{\alpha _{1}\epsilon }{6(\alpha _{1}-4\beta_1 )} ,\nonumber\\
\nu ^{-1} &=&2+\frac{5\alpha _{1}\epsilon }{3(\alpha _{1}-4\beta _{1})}%
,~~~~\! \beta /\nu =\frac{d-2+\eta}{2} .\label{exppotts}
\end{eqnarray}
They are all real and, again, return to the scalar ones if $\alpha_1=\beta_1=1$~\cite{zhongl05,zhong16}. $z$ also agrees with the same one for a $q$-state Potts model coupled with turbulent mixing~\cite{Antonov1}. It is apparent that $\beta\neq1$ from Eq.~(\ref{exppotts}). Although we have only computed the exponents to first-loop order, the three-loop results of the equilibrium exponents are available~\cite{Alcantara}.

The final ingredient of the theory is FTS, which enables one to probe into the metastable states. It is realized by cooling linearly as $\tau=-Rt$ for a constant rate $R>0$. Consider $G(\mathbf{x})$, the inverse Fourier transform of $G(\mathbf{k})$, and integrate it over the space. In equilibrium, the result is proportional to the susceptibility $\chi$. In the driven nonequilibrium case, the two functions deviate from each other near the dynamic transition point~\cite{Feng}. Still, their corresponding exponents are identical~\cite{Feng}. Consequently, we still denote the spatially integrated correlation function by $\chi$. Its FTS form can simply be found by replacing the rescaled variable for $\lambda t$ in Eq.~(\ref{form4p}) with that for $R$ and choosing a scale $\kappa$ such that $R\kappa^{-r}$ is a constant, where $r$, the RG eigenvalue of $R$, is given by~\cite{zhonge,Zhong1,Zhong2}
\begin{equation}
r=z+1/\nu.\label{rznu}
\end{equation}
The result is
\begin{equation}
\chi(\tau,R)=R^{-\gamma /r\nu
}f(\tau R ^{-1/r\nu }), \label{chifts}
\end{equation}
where the instability exponent $\gamma=d\nu-2\beta$ and $f$ is a universal scaling function.

In order to compare with the numerical results found in Ref.~\cite{Pelissetto16}, we employ some characteristic loci. These can be the peaks of $\chi$ for a series of $R$, for example. At these peaks, $\tau$ is asymptotically proportional to $R^{1/r\nu}$ from the FTS from~(\ref{chifts}). It is the hysteresis exponent $1/r\nu$ that is found to be about $1/3$ for both $q=10$ and $20$~\cite{Pelissetto16}. Note that different from Ref.~\cite{Pelissetto16} in which the temperature is measured from the equilibrium transition point and thus there is a nonzero finite temperature (or $s_2^*$ in~\cite{Pelissetto16}) at the dynamic spinodal point, our $\tau$ is directly the distance from the shifted spinodal point. To one-loop order, Eq.~(\ref{exppotts}) yields
\begin{eqnarray}
1/r\nu &=& \frac{1}{2}\left[1+\frac{3\alpha_1\epsilon}{8(\alpha_1-4\beta_1)}\right],\label{1rnu}\\
\gamma/r\nu &=& \frac{1}{2}\left[1-\frac{5\alpha_1\epsilon}{8(\alpha_1-4\beta_1)}\right].\label{grnu}
\end{eqnarray}
From Eqs.~(\ref{1rnu}) and (\ref{grnu}), together with Eq.~(\ref{qcontract}), the one-loop results are $1/r\nu=0.2$ and $0.23$ and $\gamma/r\nu=2$ and $1.9$ for $q=10$ and $20$, respectively, in $d=2$. We shall not refine these results by considering the available equilibrium exponents to the three-loop order~\cite{Alcantara}, because the dynamic exponent $z$ that composes the exponents via Eq.~(\ref{rznu}) has one-loop result only. Nevertheless, the values of the hysteresis exponent are not far away from the numerical result of about $1/3$. In contrast, the values of the exponent for the susceptibility, $\gamma/r\nu$, are different from $2/3$ found~\cite{Pelissetto16}. However, because only logarithmic factors of integer powers were considered via data collapses to find these exponents numerically~\cite{Pelissetto16}, the differences in $\gamma/r\nu$ and also in $1/r\nu$ are not quite unexpected. In addition, the results for the two $q$ values are close, in consistence with the numerical results, although the exponents all depend on $q$ at least for the $\epsilon$ expansion.

Summarizing, we have developed a dynamic RG theory for the cooling FOPTs in the Potts model. We have found that these transitions constitute the purely thermal class of driven thermal FOPTs, different from the scalar models. We have proposed that the well-known imaginary fixed points of the Potts model for $q>10/3$ in the RG analyses are the origin of the dynamic scaling found in Ref.~\cite{Pelissetto16}, apart from the logarithmic corrections. This indicates that the real and imaginary fixed points of the Potts model are both physically accessible and control the scaling of the continuous and discontinuous phase transitions, respectively, of the model for different $q$. Our one-loop results for the hysteresis scaling exponents, $0.2$ and $0.23$ for $q=10$ and $20$, respectively, are already not far away from the numerical results of about $1/3$, although the values of the exponent for the susceptibility appear different. This is not unreasonable since integer powers of logarithmic factors were employed to collapse data in numerics~\cite{Pelissetto16}. Further, the differences in the exponents for different $q$ of the scaling exponents are small, also in consistence to the numerical results. Therefore, we believe that the theory provides a natural explanation of the dynamic scaling including the scaling exponents and their scaling laws for various observables for the cooling FOPTs in the Potts model.

We thank Shuai Yin for his helpful information. This work was supported by the National Natural Science foundation of PRC (Grant No 11575297).


\end{document}